\documentclass[aps,prc,twocolumn,nofootinbib,amsmath,showpacs,superscriptaddress,floatfix]{revtex4}
\usepackage{graphicx,epsfig,longtable}
\usepackage{mathptmx}
\usepackage[light,first]{draftcopy} 
\usepackage{epstopdf}
\usepackage[pdftex]{hyperref}
\usepackage{wasysym}

\newcommand{\beqy}{\begin{eqnarray}}
\newcommand{\eeqy}{\end{eqnarray}}
\newcommand{\bmlet}{\begin{subequations}}
\newcommand{\emlet}{\end{subequations}}
\newcounter{saveeqn}

\def\gsimeq{\,\,\raise0.14em\hbox{$>$}\kern-0.76em\lower0.28em\hbox  
{$\sim$}\,\,}  
\def\lsimeq{\,\,\raise0.14em\hbox{$<$}\kern-0.76em\lower0.28em\hbox  
{$\sim$}\,\,}  

\begin{document}

\title{Fermi's golden rule applied to the $\gamma$ decay in the quasicontinuum of $^{46}$Ti}

\author{M.~Guttormsen}
\email{magne.guttormsen@fys.uio.no}
\affiliation{Department of Physics, University of Oslo, N-0316 Oslo, Norway}
\author{A.~C.~Larsen}
\affiliation{Department of Physics, University of Oslo, N-0316 Oslo, Norway}
\author{A.~B\"{u}rger}
\affiliation{Department of Physics, University of Oslo, N-0316 Oslo, Norway}
\author{A.~G{\"o}rgen}
\affiliation{Department of Physics, University of Oslo, N-0316 Oslo, Norway}
\author{S.~Harissopulos}
\affiliation{Institute of Nuclear Physics, NCSR "Demokritos", 153.10 Aghia Paraskevi, Athens, Greece}
\author{M.~Kmiecik}
\affiliation{Institute of Nuclear Physics PAN, Krak\'{o}w, Poland}
\author{T.~Konstantinopoulos}
\affiliation{Institute of Nuclear Physics, NCSR "Demokritos", 153.10 Aghia Paraskevi, Athens, Greece}
\author{M.~Krti\u{c}ka}
\affiliation{Institute of Particle and Nuclear Physics, Charles University, Prague, Czech Republic}
\author{A.~Lagoyannis}
\affiliation{Institute of Nuclear Physics, NCSR "Demokritos", 153.10 Aghia Paraskevi, Athens, Greece}
\author{T.~L\"{o}nnroth}
\affiliation{Department of Physics, \AA bo Akademi University, FIN-20500 \AA bo, Finland}
\author{K.~Mazurek}
\affiliation{Institute of Nuclear Physics PAN, Krak\'{o}w, Poland}
\author{M.~Norrby}
\affiliation{Department of Physics, \AA bo Akademi University, FIN-20500 \AA bo, Finland}
\author{H.T.~Nyhus}
\affiliation{Department of Physics, University of Oslo, N-0316 Oslo, Norway}
\author{G.~Perdikakis\footnote{Current address: National Superconducting Cyclotron Laboratory, Michigan State University, East Lansing, MI 48824-1321, USA.}}
\affiliation{Institute of Nuclear Physics, NCSR "Demokritos", 153.10 Aghia Paraskevi, Athens, Greece}
\author{A.~Schiller}
\affiliation{Department of Physics and Astronomy, Ohio University, Athens, Ohio 45701, USA}
\author{S.~Siem}
\affiliation{Department of Physics, University of Oslo, N-0316 Oslo, Norway}
\author{A.~Spyrou\footnotemark[\value{footnote}]}
\affiliation{Institute of Nuclear Physics, NCSR "Demokritos", 153.10 Aghia Paraskevi, Athens, Greece}
\author{N.U.H.~Syed}
\affiliation{Department of Physics, University of Oslo, N-0316 Oslo, Norway}
\author{H.K.~Toft}
\affiliation{Department of Physics, University of Oslo, N-0316 Oslo, Norway}
\author{G.M.~Tveten}
\affiliation{Department of Physics, University of Oslo, N-0316 Oslo, Norway}
\author{A.~Voinov}
\affiliation{Department of Physics and Astronomy, Ohio University, Athens, Ohio 45701, USA}

\date{\today}

\begin{abstract}
Particle-$\gamma$ coincidences from the $^{46}$Ti($p,p' \gamma$)$^{46}$Ti inelastic scattering reaction with 15-MeV protons are utilized to obtain $\gamma$-ray spectra as a function of excitation energy. The rich data set allows analysis of the coincidence data with various gates on excitation energy. For many independent data sets, this enables a simultaneous extraction of level density and radiative strength function (RSF). The results are consistent with one common level density. The data seem to exhibit a universal RSF as the deduced RSFs from different excitation energies show only small fluctuations provided that only excitation energies above $3$~MeV are taken into account. If transitions to well-separated low-energy levels are included, the deduced RSF may change by a factor of $2-3$, which might be expected because the involved Porter-Thomas fluctuations.
\end{abstract}


\maketitle

\section{Introduction}
\label{sec:int}
Fermi's golden rule predicts the transition rate from one state to a set of final states in a quantum system. The theoretical foundation, which has been successfully applied in many disciplines of physics, was first established by Dirac in 1927~\cite{dirac} and emphasized by Fermi in his book in 1950~\cite{fermi}. The rule is based on first-order perturbation theory, where the transition matrix element is assumed to be small. In nuclear physics, this assumption is well fulfilled for $\beta$ and $\gamma$ decay. Thus, we take the validity of the Fermi's golden rule as granted, rather than testing this rule. In this work, we study the $\gamma$ decay between states in the quasi-continuum of $^{46}$Ti and apply Fermi's golden rule to disentangle the $\gamma$ strength and level density. 

The Oslo nuclear physics group has developed a method to determine simultaneously the level density and the radiative strength function (RSF) from particle-$\gamma$ coincidences. These quantities provide information on the average properties of excited nuclei and are indispensible in nuclear reaction theories as they are the only quantities needed for complete description of the $\gamma$ decay at higher excitation energies. 

The nuclear level density can be determined reliably up to a few MeV of excitation energy from the counting of low-lying discrete known levels~\cite{ENSDF}. There is also reliable level-density information from neutron resonances at higher excitation energies; however, this information is restricted in energy as well as spin range. 

The most rich experimental information on the RSF was obtained from the study of photonuclear cross-sections~\cite{dietrich}, and thus limited only to energies above the particle separation energy. It was established that the giant electric dipole resonance (GEDR) dominates the RSF in all nuclei. The information on RSF below particle separation energy is significantly less. It has been obtained mainly from the Oslo method and (n,$\gamma$) and ($\gamma$,$\gamma^\prime$) reactions.

The Oslo method, which is applicable for excitation energies below the particle separation, is described in detail in Ref.~\cite{Schiller00}. In this work, we report on results obtained for the $^{46}$Ti nucleus, which has two protons and four neutrons outside the doubly-magic $^{40}$Ca core. The advantages of the $(p,p')$ reaction compared to the commonly used ($^3$He,$^3$He$^\prime$) and ($^3$He,$^4$He) reactions are much higher cross-sections and better particle resolution. This allows us to make a detailed study of $\gamma$-decay as a function of excitation energy.

It can be discussed if the concept of one unique RSF is valid for light nuclei and in particular at low excitation energies. This question together with the applicability of the Oslo method for light systems as titanium is the main subject of this work. 

In Sec.~II the experimental results are described. The nuclear level density and RSF are extracted in Sec.~III, and in Sec.~IV, the application of Fermi's golden rule and the Brink hypothesis is discussed. Summary and conclusions are given in Sec.~V.

\section{Experimental results}
\label{sec:exp}

The experiment was conducted at the Oslo Cyclotron Laboratory (OCL) with a 15-MeV proton beam bombarding a self-supporting target of $^{46}$Ti with thickness of 1.8 mg/cm$^2$. The target was enriched to 86\% $^{46}$Ti with $^{48}$Ti (11\%) as the main impurity. This small admixture of $^{48}$Ti is not expected to play a major role. This is supported by the fact that at low excitation energies where the level density is small, we could not identify any important contributions of $^{48}$Ti into the $\gamma$-spectra. In addition, both titaniums are expected to behave similarly.

\begin{figure}[bt]
 \begin{center}
 \includegraphics[clip,width=\columnwidth]{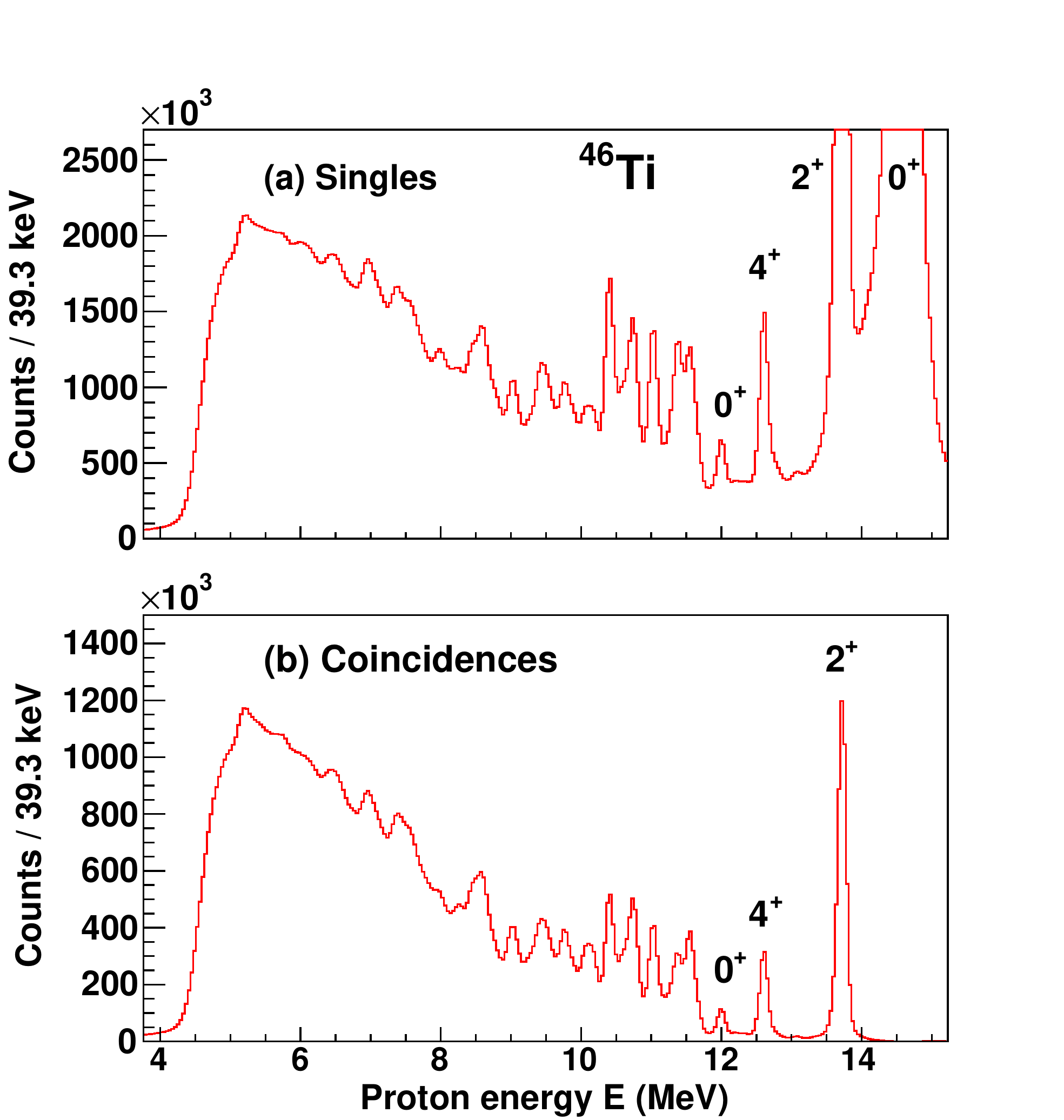}
 \caption{(Color online) Singles (a) and coincidence (b) proton spectra recorded with 15-MeV protons on $^{46}$Ti.}
 \label{fig:sing_coin}
 \end{center}
 \end{figure}

Particle-$\gamma$ coincidences were measured with eight Si $\Delta E - E$ particle telescopes and the CACTUS multidetector system~\cite{CACTUS}. The Si detectors were placed in forward direction, $45^\circ$ relative to the beam axis. The front ($\Delta E$) and end ($E$) detectors had a thickness of $140$~$\mu$m and $1500$~$\mu$m, respectively. The CACTUS array consists of 28 collimated $5" \times 5"$ NaI(Tl) $\gamma$ detectors with a total efficiency of $15.2$\% at $E_\gamma = 1.33$ MeV.  
The singles-proton spectrum and protons in coincidence with $\gamma$-rays are shown in Fig.~\ref{fig:sing_coin}. 

\begin{figure}[bt]
 \begin{center}
 \includegraphics[clip,width=\columnwidth]{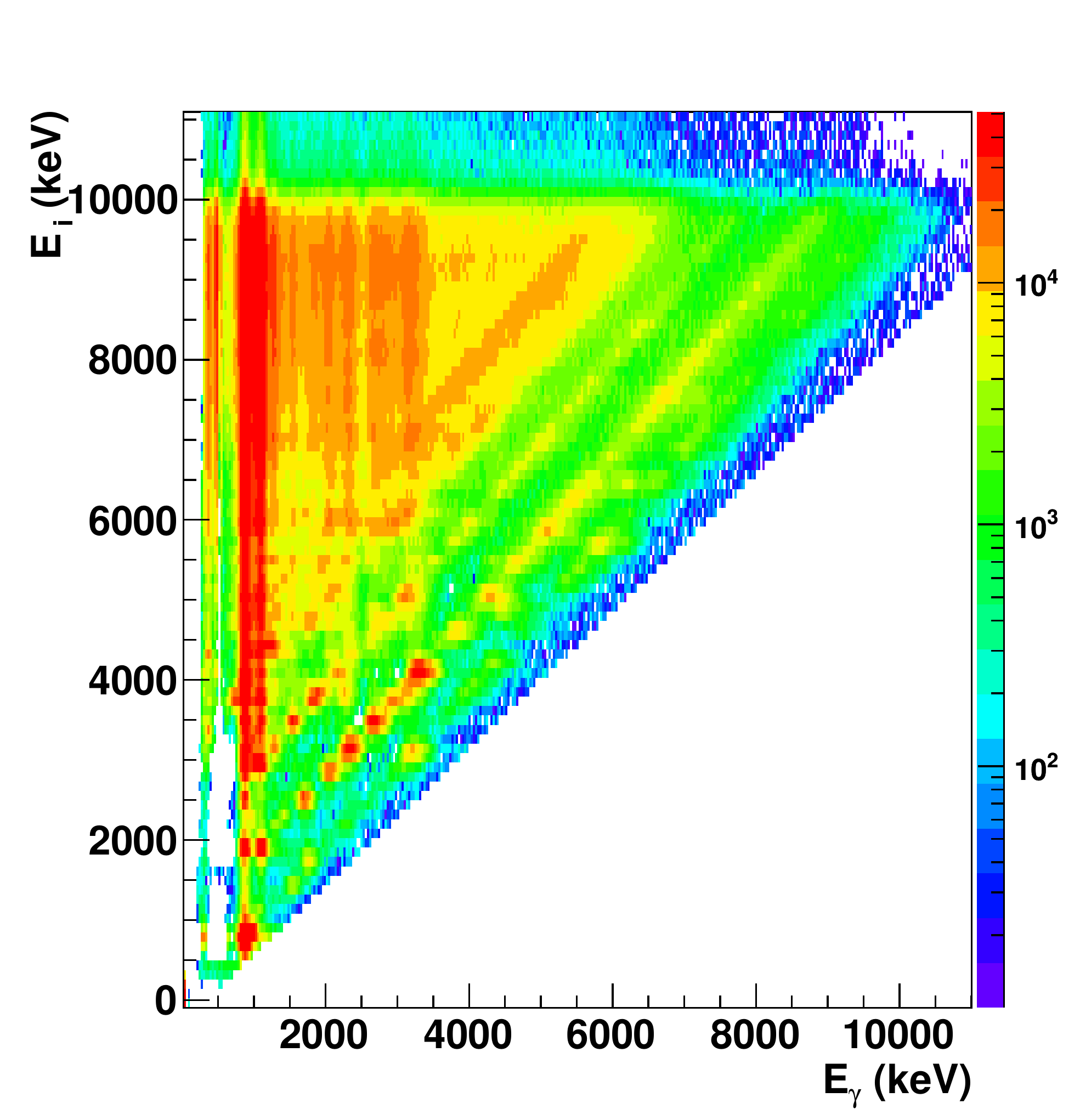}
 \caption{(Color online) The particle-$\gamma$ coincidence matrix for $^{46}$Ti. The $\gamma$-ray spectra have been unfolded with the NaI response functions.}
 \label{fig:alfnaun}
 \end{center}
 \end{figure}

In total, 110 million coincidence events were collected in one week with a beam current of 1.5 nA. Using reaction kinematics, the measured proton energy was transformed into excitation energy of the residual nucleus. In this way, a set of $\gamma$-ray spectra is assigned to a specific initial excitation energy $E_i$ in $^{46}$Ti. Furthermore, the $\gamma$-ray spectra are corrected for the known response functions of the CACTUS array following the procedure described in Ref.~\cite{gutt6}. The unfolded coincidence matrix $(E_i,E_\gamma)$ of $^{46}$Ti is shown in Fig.~\ref{fig:alfnaun}. 

The coincidence matrix displays vertical lines that represent yrast transitions from the last steps in the $\gamma$-cascades. However, there are also clear diagonal lines where the $\gamma$ energy matches the direct decay to the ground state ($E_i=E_\gamma$) and to the first and second excited states at 889 keV ($2^+$) and 2010 keV ($4^+$), respectively. These $\gamma$-rays are primary transitions in the $\gamma$-cascades. By studying the energy distribution of all primary $\gamma$-rays originating from various excitation energies, information on the level density and RSF can be extracted simultaneously.

An iterative subtraction technique has been developed to separate out the first-generation (primary) $\gamma$-transitions from the total cascade~\cite{Gut87}. The subtraction technique is based on the assumption that the decay pattern is the same whether the levels were initiated directly by the nuclear reaction or by $\gamma$ decay from higher-lying states. This assumption is automatically fulfilled when states have the same relative probability to be populated by the two processes, since $\gamma$-branching ratios are properties of the levels themselves. If the excitation bins contain many levels, it is likely to find the same $\gamma$-energy distribution from this set of levels independent of the type of population. However, the assumption is more problematic if the decay involves only a few (but not one) levels within the energy bin.

Fermi's golden rule predicts that the decay probability may be factorized into the transition matrix element between the initial and final states, and the density at the final state~\cite{fermi}: 
\begin{equation}
\lambda_{i \rightarrow f} =   \frac{2 \pi}{\hbar} |{\langle}f |H^{\prime}|i {\rangle} | ^2  {\rho}_f.\
\label{eqn:1}
\end{equation}

Realizing that the first generation $\gamma$-ray spectra $P(E_i,E_{\gamma})$ are proportional to the decay probability from $E_i \rightarrow E_f = E_i - E_{\gamma}$, i.e.~$\lambda_{i \rightarrow f} $, we may write the equivalent expression of Eq.~(\ref{eqn:1}) as: 
\begin{equation}
P(E_i, E_{\gamma}) \propto   {\cal{T}}_{i \rightarrow f}  \rho_f,\
\label{eqn:2}
\end{equation}
where ${\cal {T}}_{i \rightarrow f}$ is the $\gamma$-ray transmission coefficient, and $\rho_f = \rho (E_i -E_{\gamma})$ is the level density at the excitation energy $E_f$ after the primary $\gamma$-ray emission. This expression does not allow us to extract simultaneously ${\cal{T}}_{i \rightarrow f}$ and $\rho $ from the experimental $P(E_i, E_{\gamma})$ matrix. To do that, either one of the factorial functions must be known, or some restrictions have to be introduced. According to the Brink hypothesis~\cite{brink}, the $\gamma$-ray transmission coefficient is independent of excitation energy; only the transitional energy $E_{\gamma}$ plays a role. Thus, we replace ${\cal {T}}_{i \rightarrow f}$ with ${\cal {T}}(E_{\gamma})$, giving
\begin{equation}
P(E_i, E_{\gamma}) \propto   {\cal{T}}(E_{\gamma}) \rho_f,\
\label{eqn:3}
\end{equation}
which permits a simultaneous extraction of the two multiplicative functions.
We then fit about $N^2/2$ data points of the $P$ matrix with $2N$ free parameters. A least $\chi ^2$ fit is then possible, since it is many more data points than fit parameters; in the present case we have 150 free parameters to fit 2240 data points.

 \begin{figure*}[hbt]
 \begin{center}
 \includegraphics[clip,totalheight=8cm]{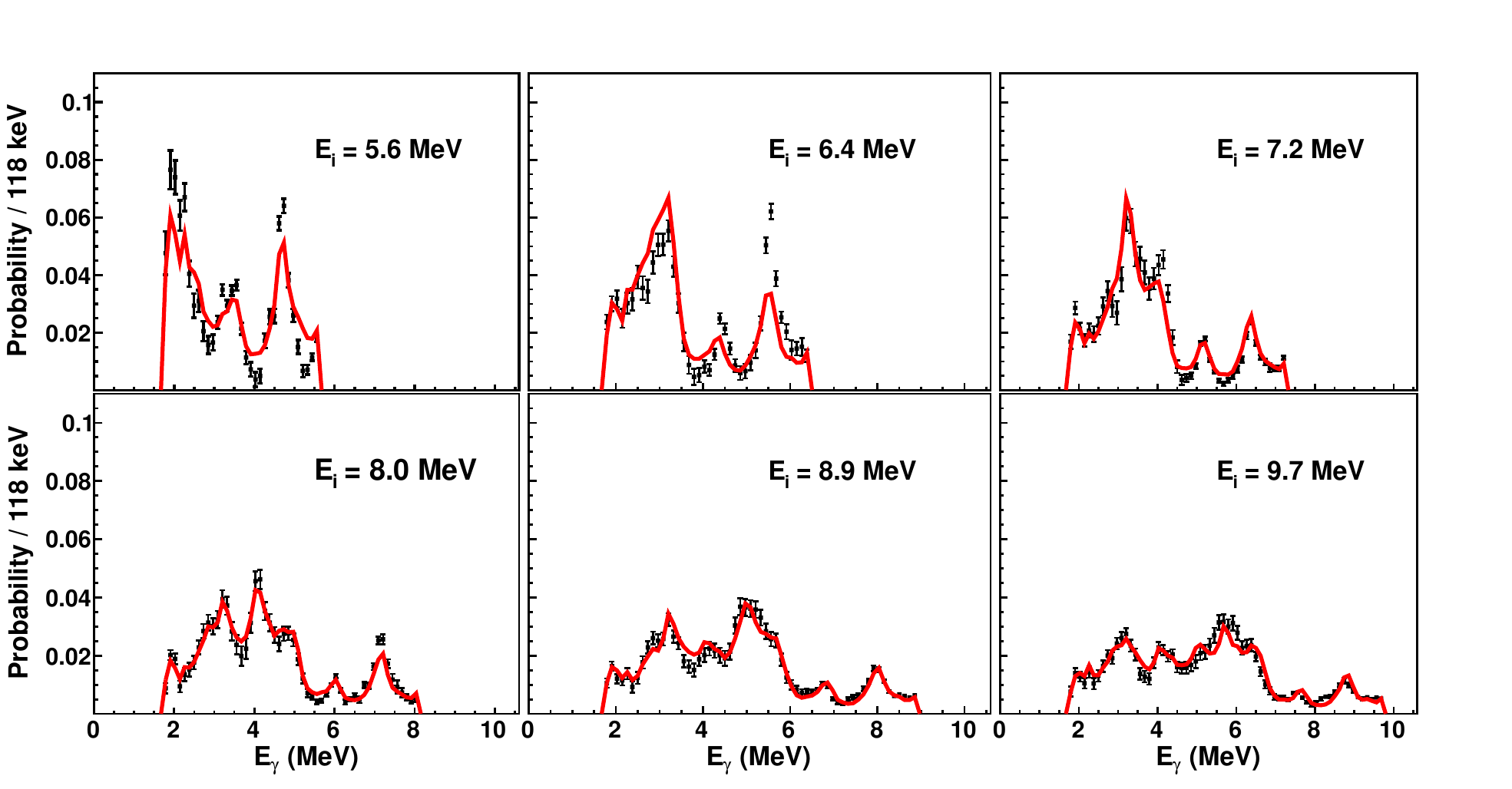}
 \caption{(Color online) Comparison of experimental first-generation spectra (squares) and the ones obtained from multiplying the extracted $\cal{T}$ and $\rho$ functions (red line). The initial excitation energy bins $E_i$ are 118 keV broad. The error bars represent the experimental statistical errors.}
 \label{fig:work}
 \end{center}
 \end{figure*}

At low excitation energy, the $\gamma$ decay is highly dependent on the individual initial and final state; therefore, we have excluded $\gamma$-ray spectra originating from excitation energy bins below $E_i = 5.5$ MeV. 

It is well known that the Brink hypothesis is violated when high temperatures and/or spins are involved, see Ref.~\cite{Andreas&Thoennessen} and references therein. However, in the present Oslo experiment, the temperature reached is low ($T \sim 1.5$ MeV) and is assumed to be rather constant. The dependency on spin is of minor importance. The measured ratios of the $\gamma$ feeding into the ground band indicate a low spin window of $I \sim 0 - 6 \hbar$. At excitation energy $E \sim 8 $~MeV the ratios are 57:100:9 for the $2^+$, $4^+$ and $6^+$ states, and at $E \sim 10$~MeV they are 55:100:10, which are equal within the error bars for extracting these ratios. Of course, at low excitation energy the spin distribution fluctuates due to a few levels (and spins) present within each 118~keV excitation bin.

In principle, the Brink hypothesis assumed in expression~(\ref{eqn:3}) could bias the analysis, so that the Oslo method in the next turn validates the Brink hypothesis. This issue will be addressed in Sec.~IV, where we find one common level density (according to Fermi's golden rule), which in turn results in one universal RSF in the quasi-continuum of $^{46}$Ti.

\section{Level density and radiative strength function}

In our first investigations, we rely on the Brink hypothesis from the expression~(\ref{eqn:3}) and factorize the first-generation $\gamma$-matrix $P(E_i, E_{\gamma})$ into one transmission coefficient ${\cal {T}}(E_{\gamma})$ and one level density $\rho(E)$. Since the decay between states at low excitation energy cannot be treated within a statistical approach, a cut of the matrix with $E_{\gamma} > 1.8$ MeV and $5.5$ MeV $< E_i < 10.0$ MeV was used to exclude clear non-statistical decay routes\footnote{The lower excitation cut concerns only the initial excitation energy $E_i$; one still needs $\gamma$-spectra originating from excitation regions down to the ground state in order to subtract higher-order generations of $\gamma$-rays.}. The least $\chi^2$ fitting of the two multiplicative functions follows the iterative procedure of Ref.~\cite{Schiller00}.  

To demonstrate how well the procedure works, we compare in Fig.~\ref{fig:work} for some initial excitation energies $E_i$ the experimental first-generation spectra $P$ with the ones obtained by multiplying the extracted $\cal{T}$ and $\rho$ functions. The agreement between calculated and experimental first-generation spectra is excellent for decay from higher excitations; however there are locally strong deviations where the calculated spectra fall outside of the experimental error bars for populations of lower excitations energies, as seen in the $E_i = 5.6$ and $6.4$~MeV gates. These deviations could be the consequence of Porter-Thomas fluctuations~\cite{PT}, which will be further discussed in Sec.~IV. The general good agreement holds also for all the other 40 spectra (not shown) included in the global fit with the same ${\cal {T}}(E_{\gamma})$ and $\rho(E)$ functions.

The experimental statistical errors are very small as seen in Fig.~\ref{fig:work}. Thus, the deviations are due to other sources of errors. For example for $E_i=6.4$~MeV, the $\cal{T}\cdot \rho$ prediction overestimates around $E_\gamma = 3$~MeV and underestimates around $E_\gamma = 5.5$~MeV with several standard deviations. Thus, there are indications that a common $\cal{T}$ and $\rho$ function that simultaneously fits the $P(E_i,E_{\gamma})$ matrix, could not be found. The systematic errors behind these deviations could be due to several factors as experimental shortcomings, assumptions behind the Oslo method, the Brink hypothesis \cite{brink} and, most probable, Porter-Thomas fluctuations. Keeping all these possibilities in mind, the results of Fig.~\ref{fig:work} are very gratifying.

 \begin{figure}[hbt]
 \begin{center}
 \includegraphics[clip,width=\columnwidth]{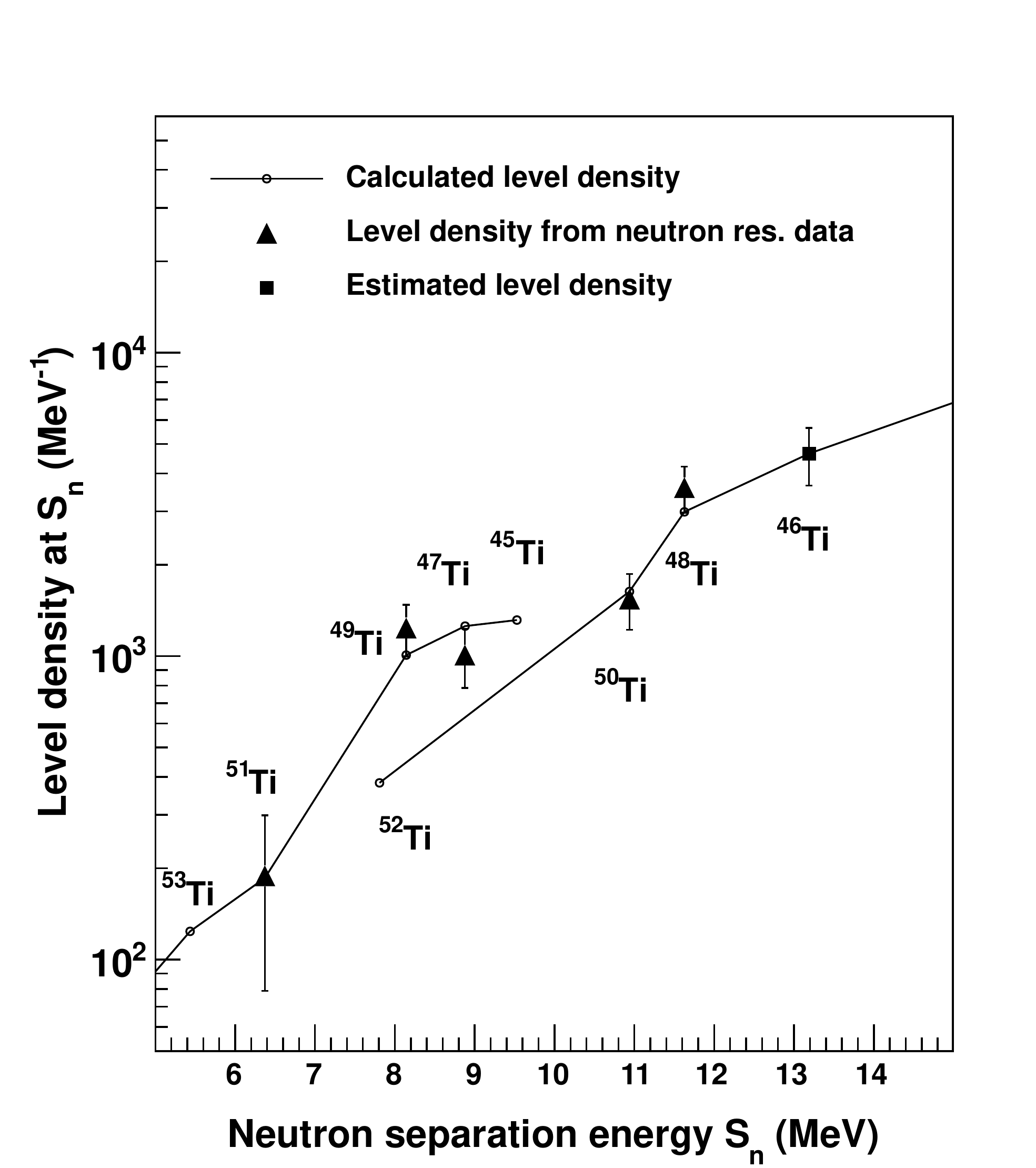}
 \caption{Deduced total $\rho(S_n)$ from neutron resonance spacings (triangles). The data point for $^{46}$Ti (square) is estimated by extrapolations from the BSFG model with global parameterization of von Egidy and Bucurescu~\cite{egidy2} (cirles with lines to guide the eye).}
 \label{fig:sys_rho}
 \end{center}
 \end{figure}

 \begin{figure}[hbt]
 \begin{center}
 \includegraphics[clip,width=\columnwidth]{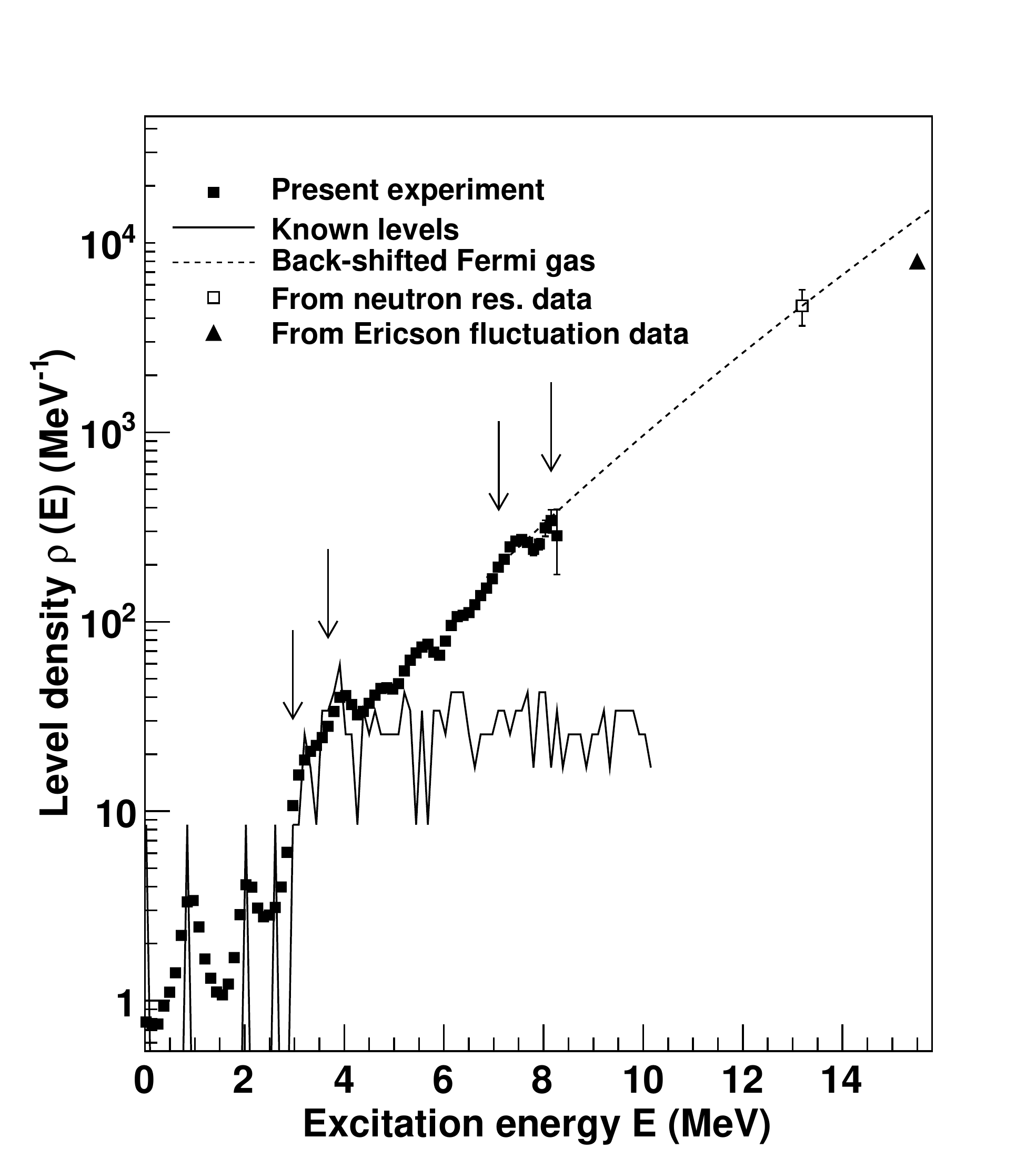}
 \caption{Normalization of the nuclear level density (filled squares) of $^{46}$Ti. At low excitation energies, the data are normalized (between the
arrows) to known discrete levels (solid line). At higher excitation energies, the data are normalized to the BSFG level density (dashed line) going through 
the point $\rho(S_n)$ (open square), which is estimated from the systematics of Fig.~\ref{fig:sys_rho}. For comparison a data point \cite{Ohio04} obtained from Ericson fluctuations are shown (black triangle).}
 \label{fig:counting}
 \end{center}
 \end{figure}

There exist infinitely many $\cal{T}$ and $\rho$ functions making identical fits to the data~\cite{Schiller00} as the examples shown in Fig.~\ref{fig:work}. These functions can be generated by the transformations 
\begin{eqnarray}
\tilde{\rho}(E_i-E_\gamma)&=&A\exp[\alpha(E_i-E_\gamma)]\,\rho(E_i-E_\gamma),
\label{eq:array1}\\
\tilde{{\mathcal{T}}}(E_\gamma)&=&B\exp(\alpha E_\gamma){\mathcal{T}} (E_\gamma).
\label{eq:array2}
\end{eqnarray}
In the following, we will try to determine the parameters $A$, $\alpha$, and $B$. This information is not available from our experiment and has to be determined from other experimental results.

First, the level density at high excitation energy is normalized to the level density at the neutron separation energy $\rho(S_n)$. This data point is calculated from neutron resonance spacings $D_0$ (see, e.g., Ref.~\cite{Pb}) with a spin distribution given by~\cite{GC} 
\begin{equation}
g(E,I) \simeq \frac{2I+1}{2\sigma^2}\exp\left[-(I+1/2)^2/2\sigma^2\right],
\label{eq:spindist}
\end{equation}
where $E$ is excitation energy and $I$ is spin.

There exists no neutron resonance data for $^{46}$Ti. We therefore estimate $\rho(S_n)$ from the parameterizations of von Egidy and Bucurescu~\cite{egidy2} using the back-shifted Fermi gas (BSFG) model,
which reads
\begin{equation}
\rho_{\rm BSFG}(E) = \eta \frac{\exp(2\sqrt{aU})}{12\sqrt{2}a^{1/4}U^{5/4}\sigma},
\label{eqn:6}
\end{equation}
where $a$ is the level density parameter, $U=E-E1$ is the intrinsic excitation energy, and $E1$ is the back-shift energy parameter. The spin cut-off parameter $\sigma$ is given by~\cite{egidy2}
\begin{equation}
\sigma^2 = 0.0146 A^{5/3} \frac{1+\sqrt{1+4aU}}{2a},
\label{eqn:8}
\end{equation}
$A$ being the nuclear mass number. 

Figure \ref{fig:sys_rho} shows the extracted total level densities for titanium isotopes with known resonance spacings $D_0$ at $S_n$ (filled triangles). The resonance spacings for $\ell=0$ neutrons only give the level densities for one or two spins and only one parity. In order to extract the level density for all spin and parities, we use Eq.~(\ref{eq:spindist}) and assume equally many positive and negative parity states at $S_n$. This is supported by combinatorial quasiparticle models \cite{ goriely,Kristine10,Naeem09}, where the numbers of positive and negative parity states at $S_n$ are predicted to be the same.

The points connected with lines are based on the semi-empirical estimate of von Egidy and Bucurescu~\cite{egidy2} with a common scaling of $\eta = 0.5$ in order to match qualitatively the experimental $\rho(S_n)$ points. The estimated value for $^{46}$Ti is $\rho= (4650 \pm 1000)$~MeV$^{-1}$ at $S_n=13.189$~MeV (see Fig.~\ref{fig:sys_rho}). The error bar chosen reflects roughly the general deviation between the global estimates and the points derived from neutron resonance data. 

Now the scaling ($A$) and slope ($\alpha$) parameters of the level density can be determined as shown in Fig.~\ref{fig:counting}. The normalization of the level density is fitted to the known level density around 3.5 MeV of excitation energy, and to the extrapolation from $\rho(S_n)$ using the BSFG model with parameters summarized in Table~\ref{tab:parameters}. By choosing other $\rho(S_n)$ values (within the assumed uncertainty of $\pm1000$~MeV$^{-1}$), the logarithmical slope of the $\rho$ and $\mathcal{T}$ will change accordingly, as the $\alpha$-parameter of Eqs.~(\ref{eq:array1}, \ref{eq:array2}) has to be adjusted. The $\rho$ curve is well fixed at $\sim 3.5$ MeV and will rotate around this point with different choices of $\alpha$. Level densities in the $14-19$ MeV excitation region of $^{46}$Ti have been measured from Ericson fluctuations, see Ref.~\cite{Ohio04} and references therein. However, the values reported by the various experimental groups differ within a factor of ten. The measurement of the Ohio group \cite{Ohio04} is shown in Fig.~\ref{fig:counting} for comparison (see triangle point at 15.5 MeV).

One should stress that the level density found from the experiment is based on the spin and parity levels populated in the $(p, p')$ reaction. Thus, the normalization to the total level density described above, rests on the assumption that the structure of the level density remains approximately the same if all spins and parities are included. This assumption is reasonable fulfilled according to Ref.~\cite{Kristine10}, where the level density for spin windows of $2-6\hbar $ and $0-30 \hbar$ have been calculated for $^{46}$Ti within a combinatorial quasiparticle model~\cite{Naeem09}. From these estimates, our measurement includes $70-80$\% of the total level density and the level density fine structures for the two spin windows are very similar. 

The measured level density describes nicely the known level density up to around 4 MeV of excitation energy. The experimental resolution of $\rho$ is about $0.3$~MeV at low excitation energy, as seen for the $2^+$ state at 889 keV. At around 3 MeV, we see an abrupt increase in level density due to the breaking of proton and/or neutron pairs coupled in time-reversed orbitals (Cooper pairs \cite{BCS}). This results in $\sim 10$ times more levels, making it difficult to determine the level density at higher excitation energies with conventional spectroscopic methods.

 \begin{figure}[]
 \begin{center}
 \includegraphics[clip,width=\columnwidth]{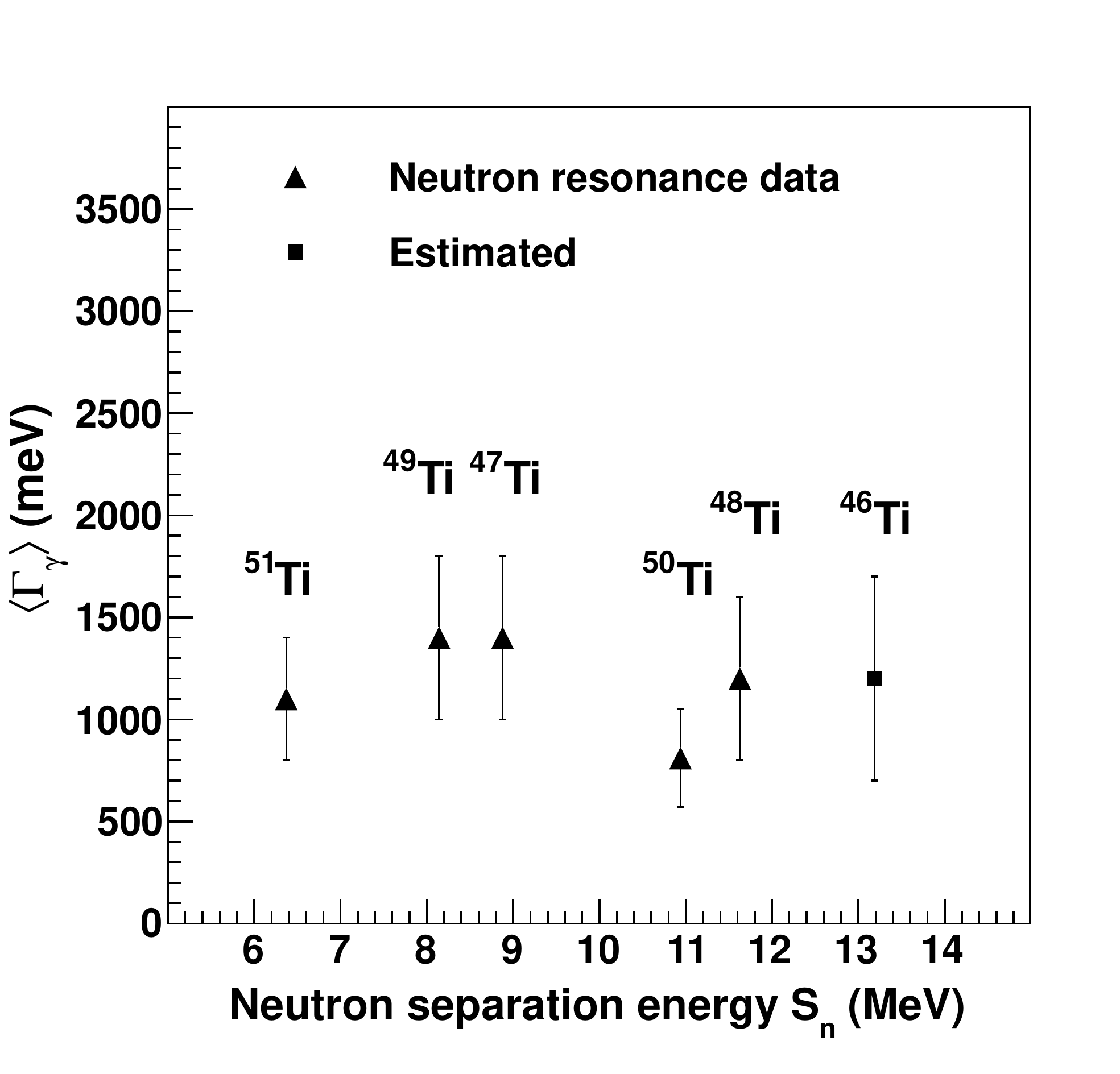}
 \caption{Experimental average $\gamma$-width $\langle \Gamma_{\gamma} \rangle$ from neutron resonance data (triangles) \cite{RIPL3}. Only a rough estimate for $^{46}$Ti can be achieved (square).}
 \label{fig:sys_gamma}
 \end{center}
 \end{figure}

 \begin{figure}[hbt]
 \begin{center}
 \includegraphics[clip,width=\columnwidth]{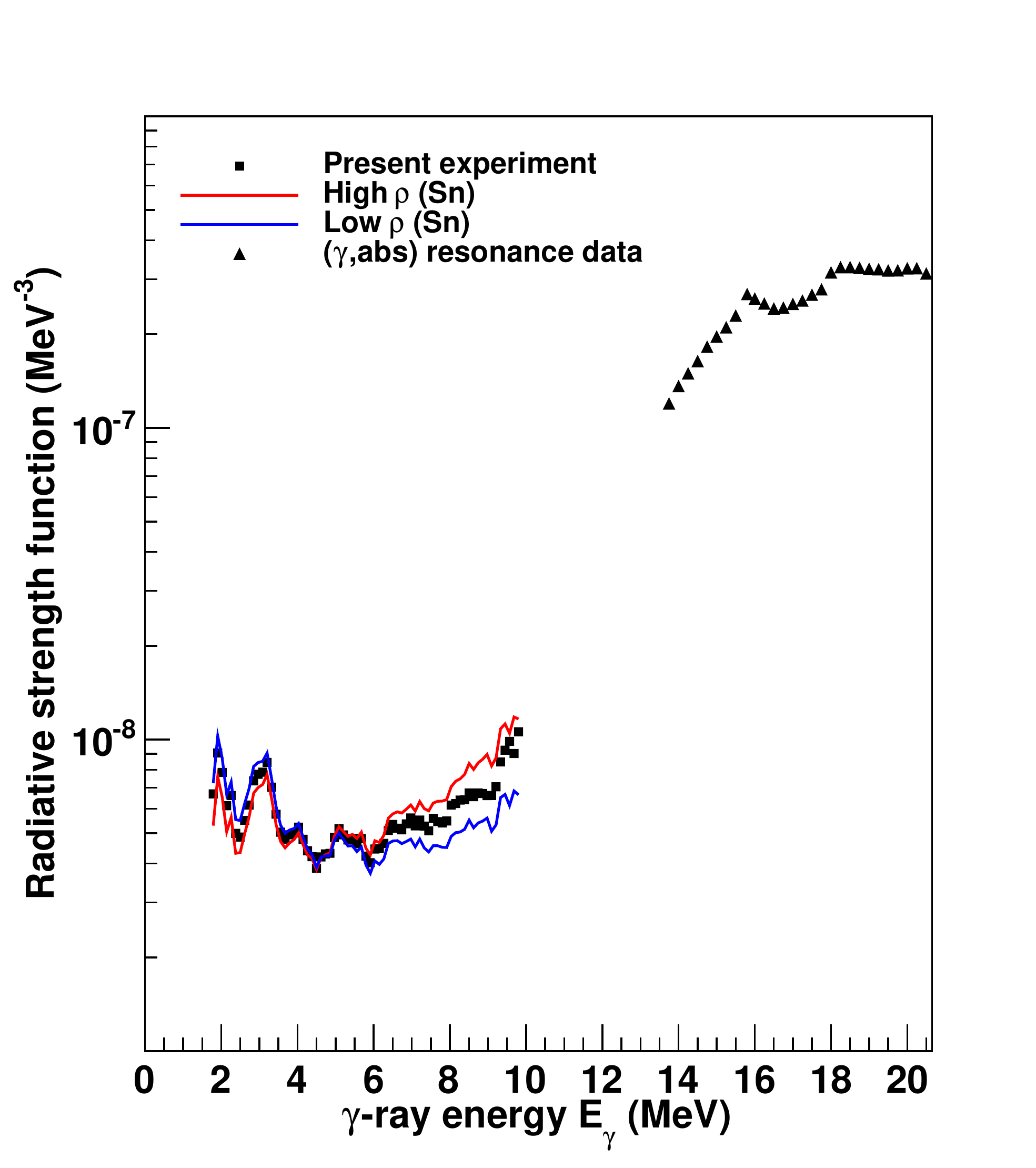}
 \caption{(Color online) Experimental radiative strength function for $^{46}$Ti (squares). The RSFs assuming high (5650 MeV$^{-1}$) and low (3650 MeV$^{-1}$) level density at $S_n$ are also shown. For comparison the GEDR data from the ($\gamma,{\rm abs}$) reaction \cite{Ishkhanov} are shown (triangles).}
 \label{fig:strength}
 \end{center}
 \end{figure}

It remains to determine the scaling parameter $B$ of the transmission coefficient ${\cal {T}}(E_{\gamma})$. Here we use the radiative width $\langle \Gamma_{\gamma} \rangle$ at $S_n$ assuming that the $\gamma$-decay is dominated by dipole transitions. For initial spin $I$ and parity $\pi$, the width is given by~\cite{ko90} 
\begin{eqnarray}
\langle\Gamma_\gamma\rangle=\frac{1}{2\pi\rho(S_n, I, \pi)} \sum_{I_f}&&\int_0^{S_n}{\mathrm{d}}E_{\gamma} B{\mathcal{T}}(E_{\gamma})
\nonumber\\
&&\times \rho(S_n-E_{\gamma}, I_f),
\label{eq:norm}
\end{eqnarray}
where the summation and integration run over all final levels with spin $I_f$ that are accessible by $\gamma$ radiation with energy $E_{\gamma}$ and multipolarity $E1$ or $M1$. Further details on the normalization procedure are found in Refs.~\cite{Schiller00,voin1}.

Since there exists no neutron resonance data, we again have to rely on systematics from other isotopes. From Fig.~\ref{fig:sys_gamma}, we estimate an average $\gamma$-width of $\langle \Gamma_{\gamma} \rangle = (1200 \pm 500)$~meV at $S_n$. The large uncertainty in $\langle \Gamma_{\gamma} \rangle$ gives an absolute normalization of ${\cal {T}}(E_{\gamma})$ and the RSF within about $\pm 50$\%. However, this uncertainty does not influence the energy dependence.

The deduced RSF for dipole radiation can be calculated from the normalized transmission coefficient ${\cal {T}}(E_{\gamma})$ by~\cite{RIPL3}
\begin{equation}
f (E_{\gamma}) =\frac{1}{2\pi} \frac{ {\mathcal{T}}(E_{\gamma})}{ E_{\gamma}^3}.
\label{eq:fT}
\end{equation}
The normalized RSF is shown in Fig.~\ref{fig:strength}. For comparison, the GEDR data from Ref.~\cite{Ishkhanov} are also shown, which have been translated from photo neutron cross section $\sigma$ to RSF by~\cite{RIPL3}
\begin{equation}
f (E_{\gamma}) =\frac{1}{3\pi^2 \hbar^2c^2} \frac{\sigma(E_{\gamma)}}{ E_{\gamma}}.
\label{eq:fT2}
\end{equation}
Unfortunately, there is a large energy gap between our data ending at a $E_{\gamma} = 10$~MeV and the GEDR data that start at $14$~MeV. 

Our data on the RSF display a minimum near $4-6$ MeV and some structures at lower $\gamma$-ray energies. The uncertainties in the value of $\rho$ at $S_n$ will change the slope of the RSF, and thus also the degree of low-energy enhancement. However, the lines of Fig.~\ref{fig:strength} show that the enhancement is not very sensitive to reasonable choices of the value $\rho(S_n)$. 

Such or similar enhancement has been seen in several light nuclei with mass $A<100$, see Ref.~\cite{sc} and references therein. There is still no theoretical explanation for this very interesting phenomenon.

\begin{table}[htb]
\caption{Parameters used for the extraction of level density and radiative strength function in $^{46}$Ti.} 
\begin{tabular}{|cccc|cc|cc|}
\hline
\hline
$S_n$&$a$&$E_1$&$\sigma$&$\rho(S_n)$&$\langle \Gamma_{\gamma}(S_n)  \rangle$ \\
(MeV)&(MeV$^{-1}$)&(MeV)& &(MeV$^{-1}$)  & (meV) \\
\hline
13.189  & 4.7 & -1.0 & 4.0 &4650(1000)&  1200(500)   \\
\hline
\hline
\end{tabular}
\\
\label{tab:parameters}
\end{table}

 \begin{figure}[bt]
 \begin{center}
 \includegraphics[clip,width=\columnwidth]{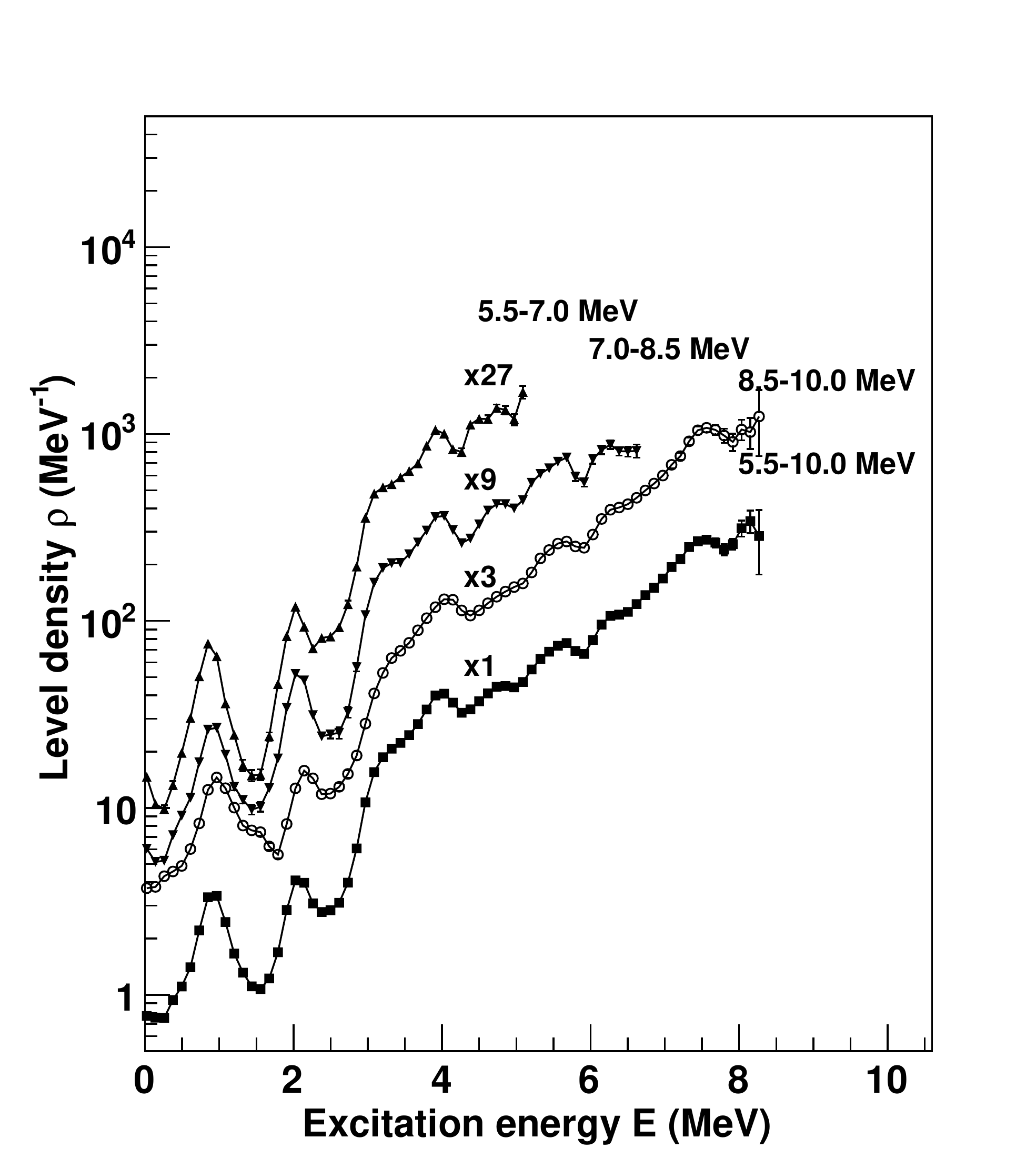}
 \caption{Level density extracted from statistically independent data sets, taken from various initial excitation energy bins $E_i$ (the three upper curves). The lower curve is the result for the whole energy region and is identical to the one of Fig.~\ref{fig:counting}.}
 \label{fig:rhos}
 \end{center}
 \end{figure}

 \begin{figure}[bt]
 \begin{center}
 \includegraphics[clip,width=\columnwidth]{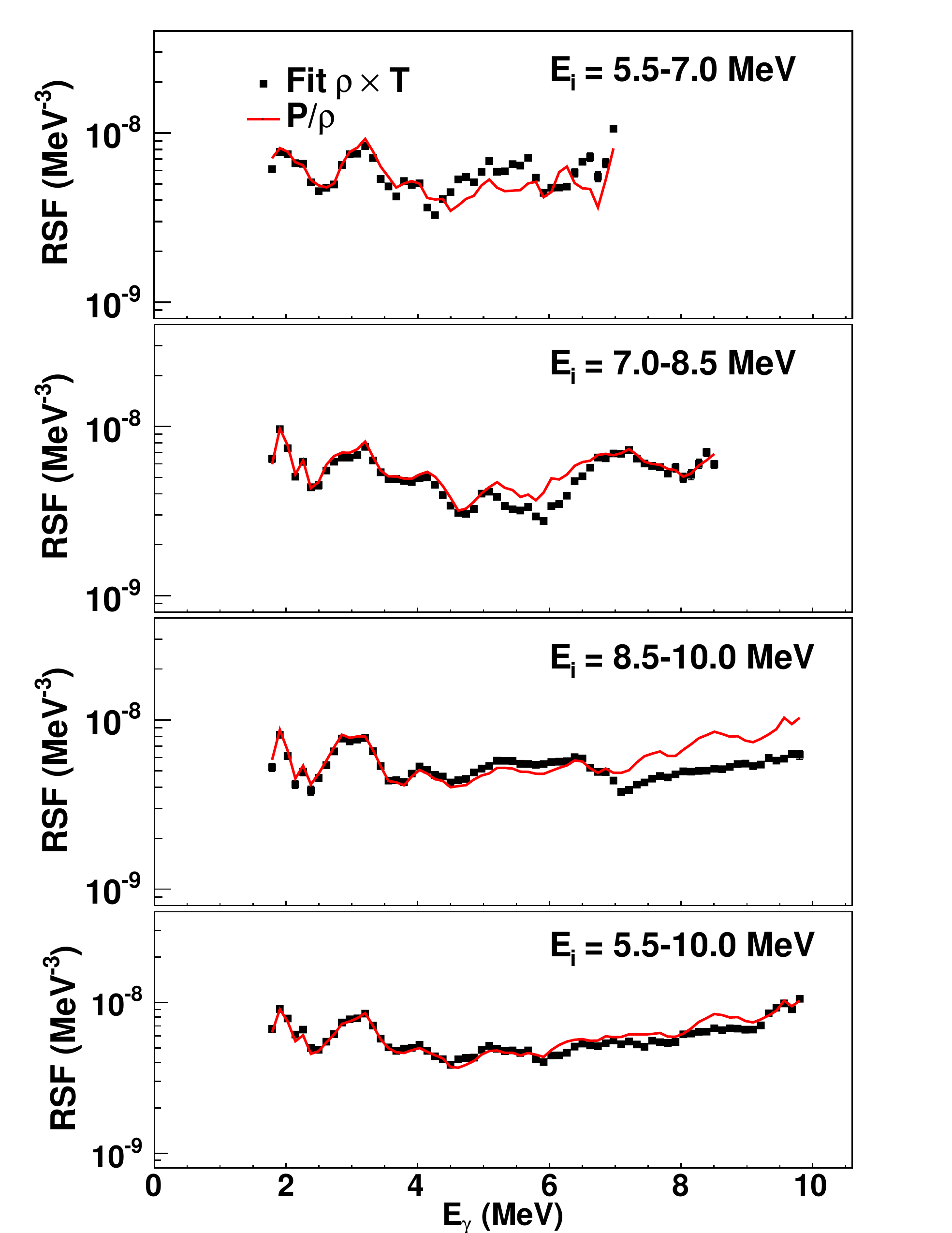}
 \caption{(Color online) Gamma-ray strength functions extracted from various initial excitation bins $E_i$. The data points of the three upper deduced RSFs are from statistically independent data sets. The data points in the lower panel are identical with the RSF of Fig.~\ref{fig:strength}. The RSFs displayed as red lines, are evaluated from the ratio $P(E_i,E_{\gamma})/\rho(E_f)$ (see text). The resemblance between the two methods confirms that the level density is common for the various excitation regions, in accordance with Fermi's golden rule.}
 \label{fig:rsf_compare}
 \end{center}
 \end{figure}

\section{Fermi's golden rule and Brink hypothesis}

According to Fermi's golden rule, it is possible to factorize out the level density from the $\gamma$-decay probability. Since the level density is a unique property of the nucleus, the same extracted level density is expected from the decay probability deduced at different excitation energy regions within the experimental restrictions, in particular the energy resolution of CACTUS. In order to see if the Oslo method gives a unique level density, we have divided the data set into three statistically independent initial excitation-energy regions, namely $E_i = 5.5 - 7.0$~MeV, $7.0 - 8.5$~MeV and $8.5 - 10.0 $~MeV, and applied the same methodology as described in Sec.~III. The results shown in Fig.~\ref{fig:rhos} are very satisfactory. The different data sets give approximately the same level density. Thus, the disentanglement of level density and transmission coefficient from $\gamma$-particle coincidences seems to work very well according to Eq.~(\ref{eqn:2}).

The fact that we measure approximately the same level density function for primary $\gamma$-ray spectra taken at different excitation regions indicates that the RSF depends only on $\gamma$-ray energy and not on excitation energy. This seems to indicate the validity of Brink hypothesis, which will be tested more thoroughly in the following. In principle, the test could be performed by dividing the data set into even more initial excitation-energy regions. However, the statistics of the experiment does not permit such an approach, and a different approach has been used.

By accepting the level density obtained with the global fit of all relevant data (see lowest curve in Fig.~\ref{fig:rhos}), we may further investigate the transmission coefficient in detail, and thus the validity of the Brink hypothesis. We first adopt the solutions ${\cal T}$ and $\rho$ from Sec.~III and write 
\begin{equation}
{\cal N}(E_i)P(E_i,E_{\gamma})\approx {\cal T}(E_{\gamma}) \rho(E_i-E_{\gamma}).
\label{eq:np}
\end{equation}
The normalization factor for each initial excitation bin is defined by
\begin{equation}
{\cal N}(E_i)=\frac{\int_0^{E_i} {\mathrm{d}} E_{\gamma } \, {\cal T}(E_{\gamma}) \rho(E_i-E_{\gamma})
}{\int_0^{E_i}{\mathrm{d}} E_{\gamma} \, P(E_i,E_{\gamma})}.
\label{eq:nei}
\end{equation}
The degree of correctness of the approximation (\ref{eq:np}) is typically illustrated by the fits in Fig.~\ref{fig:work}.

 \begin{figure*}[bt]
 \begin{center}
 \includegraphics[clip,width=1.4\columnwidth]{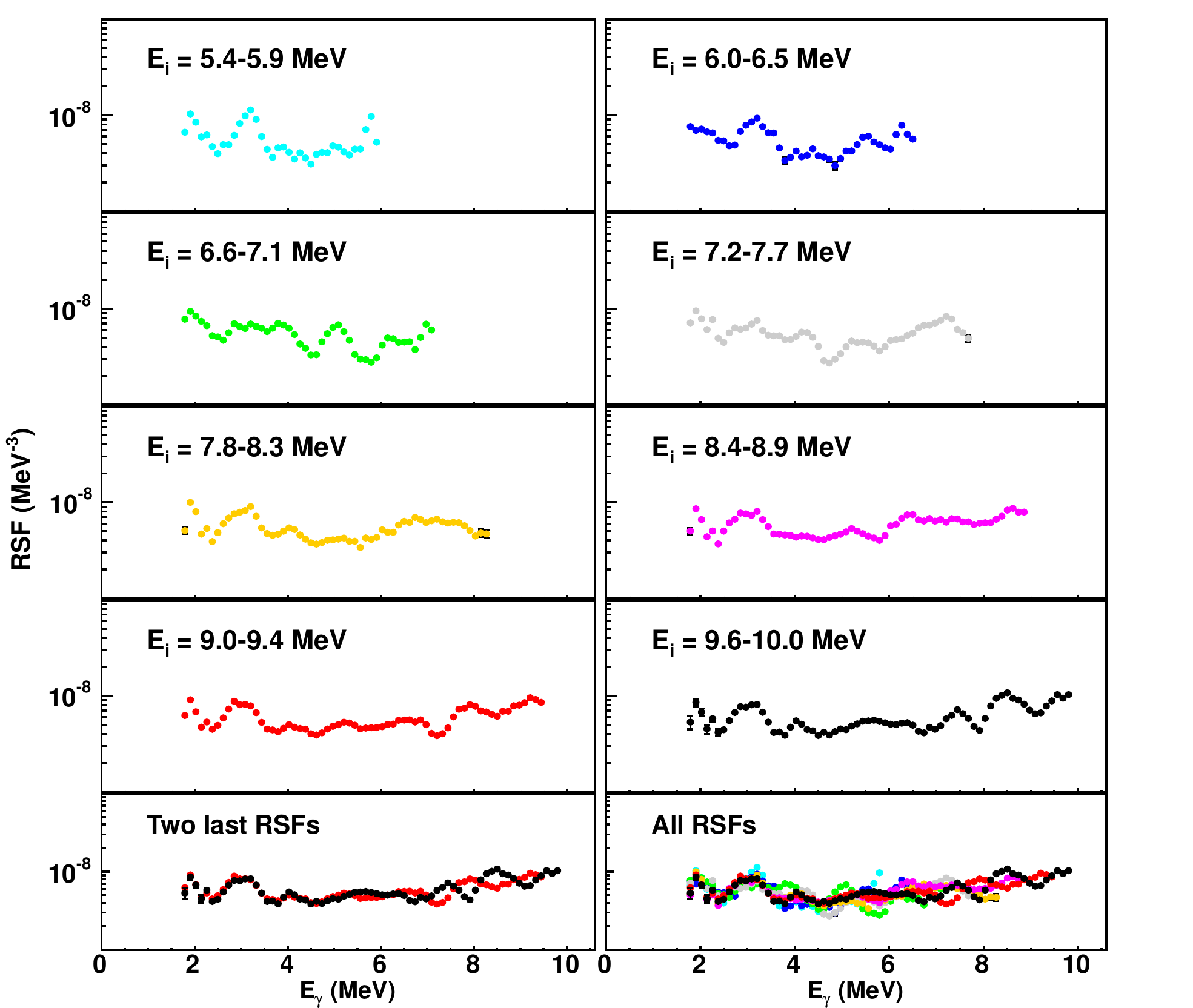}
 \caption{(Color online) Deduced RSFs from various initial excitation bins $E_i$. The RSFs are evaluated from the ratio $P(E_i,E_{\gamma})/\rho(E_f)$ as described in the text.}
 \label{fig:rsfi}
 \end{center}
 \end{figure*}

 \begin{figure*}[bt]
 \begin{center}
 \includegraphics[clip,width=1.4\columnwidth]{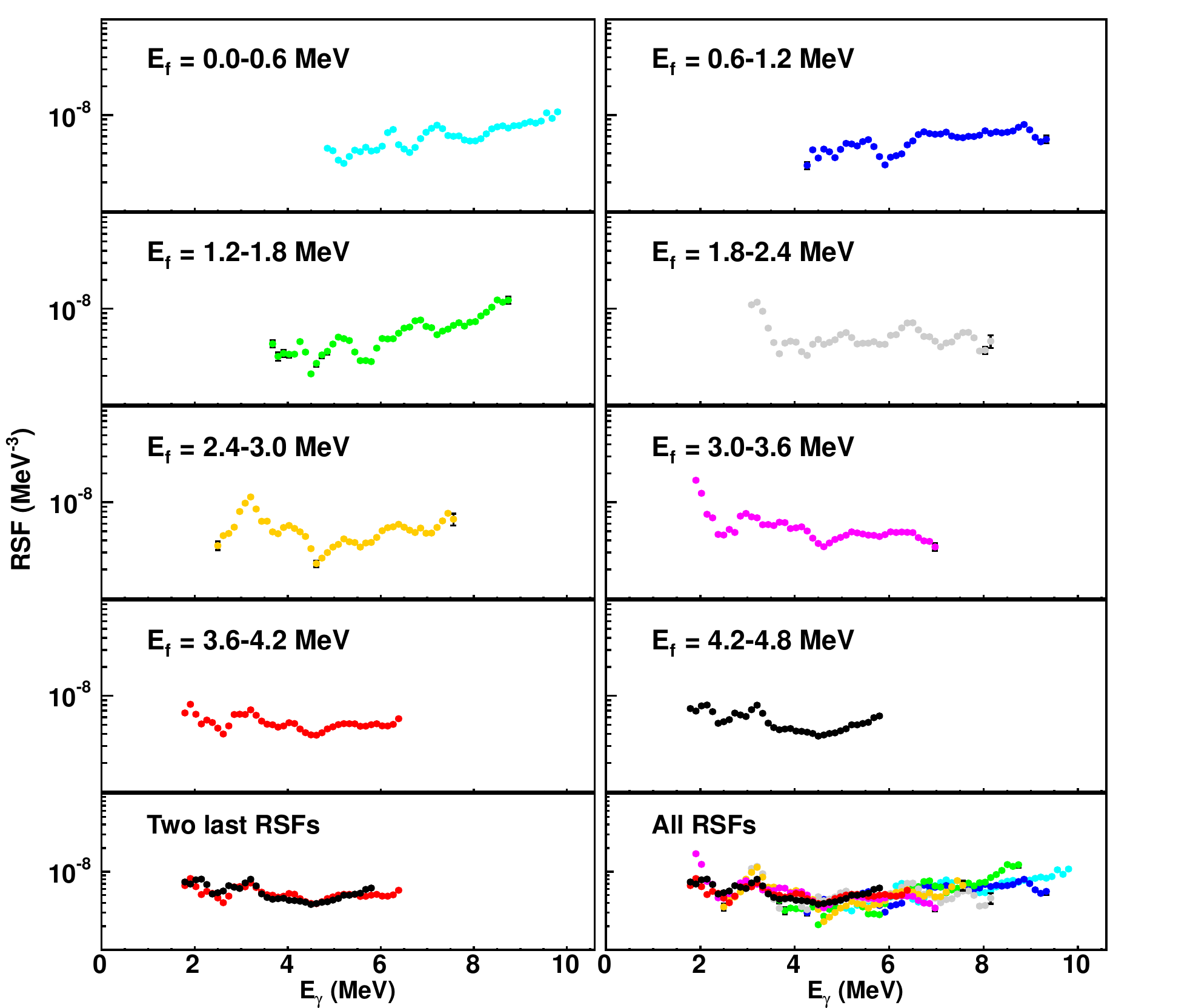}
 \caption{(Color online) Deduced RSFs populating various final excitation bins $E_f$. The RSFs are evaluated from the ratio $P(E_f+ E_{\gamma}, E_{\gamma})/\rho(E_f)$ as described in the text. It should be noted that there are no final states at $E_f= 1.2 - 1.8$~MeV, however the experimental resolution (see Fig.~\ref{fig:counting}) is responsible for including the $2^+$ and $4^+$ ground band states in this gate.}
 \label{fig:rsff}
 \end{center}
 \end{figure*}

 \begin{figure}[bt]
 \begin{center}
 \includegraphics[clip,width=\columnwidth]{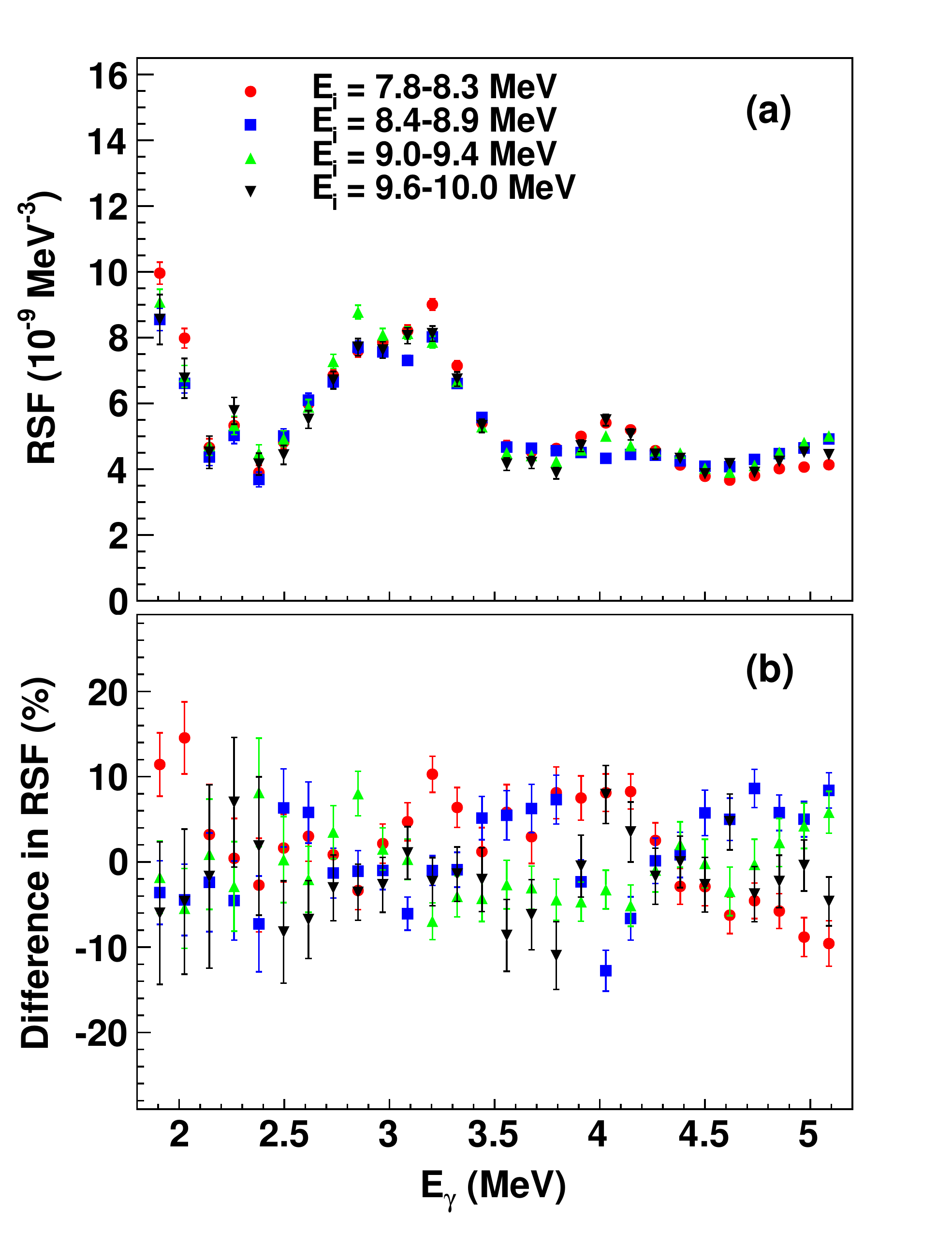}
 \caption{(Color online)(a) Radiative strength functions  for $\gamma$ transitions between states in quasi-continuum. Data from the four highest excitation energy gates of Figs.~\ref{fig:rsfi} have been chosen. (b) Ratios of the deviation from the average RSF at each $\gamma$ energy, see text.}
 \label{fig:fluct}
 \end{center}
 \end{figure}

In the following, we will investigate the dependency of the deduced RSF on initial and final excitation energy. If the Brink hypothesis is valid, there exists the same RSF for all excitation energies in this nucleus. We will call it the {\it universal} RSF. In reality the Porter-Thomas fluctuations involved influence the RSF obtained from different excitation regions. We will use the term {\it deduced} RSF in the following for the quantity obtained from experimental data. The deduced RSFs are expected to fluctuate around the universal RSF and the fluctuations are expected to be stronger if the number of transitions used in the determination of the deduced RSF is smaller. Thus, the deduced RSFs extracted from data sets involving initial and final regions of high level density should be much closer to the universal RSF than the RSFs deduced from regions of low level density.

Since there exists only one unique level density, we construct the counterpart to Eq.~(\ref{eq:np}) in the case that the transmission coefficient depends on the initial excitation energy:
\begin{equation}
{\cal N}^{\prime}(E_i)P(E_i,E_{\gamma}) \approx {\cal T}(E_{\gamma},E_i) \rho(E_i-E_{\gamma}),
\label{eq:npei}
\end{equation}
where ${\cal N}^{\prime}$ is determined analogously to Eq.~(\ref{eq:nei}). We expect that ${\cal T}(E_i, E_{\gamma})$ fluctuates on the average around ${\cal T}(E_{\gamma})$. Thus, it is reasonable to expect that ${\cal N}^{\prime} \approx {\cal N}$, which gives a transmission coefficient of:
\begin{equation}
{\cal T} (E_i, E_{\gamma}) \approx {\cal N}(E_i)
\frac{P(E_i,E_{\gamma})}{\rho(E_i - E_{\gamma})}.
\label{eq:Ei}
\end{equation}
Similarly, the transmission coefficient as a function of the final excitation energy $E_f=E_i-E_{\gamma}$ is given by 
\begin{equation}
{\cal T} (E_f, E_{\gamma}) \approx {\cal N}(E_f + E_{\gamma})
\frac{P(E_f + E_{\gamma},E_{\gamma})}{\rho(E_f)}.
\label{eq:Ef}
\end{equation}

The validity of the approximations~(\ref{eq:npei}) and (\ref{eq:Ei}) is demonstrated in Fig.~\ref{fig:rsf_compare} by comparing the deduced RSF from this approximation (lines) with the independent fits of ${\cal T}$ and $\rho$ (data points). The RSFs determined for the whole energy region $5.5-10$ MeV are very similar, especially for low $E_\gamma$, as shown in the lower panels. Also the similarity in the detailed structures of the two methods is recognized, although some differences are present. The deviations are largest for the high $\gamma$-energy part of the RSFs populating the lowest $0^+$, $2^+$, and $4^+$ states, where large Porter-Thomas fluctuations are expected (these fluctuations are not included in the error bars). However, the overall good resemblance encourages us to study the detailed evolution of the RSFs as a function of initial excitation energies using the approximations~(\ref{eq:npei}) and (\ref{eq:Ei}). Since the approximation (\ref{eq:Ef}) is a simple transformation using $E_f=E_i-E_{\gamma}$, we may also investigate the dependencies of the RSF on final excitation energy $E_f$.

In Figs.~\ref{fig:rsfi} and \ref{fig:rsff}, the RSFs $f(E_i, E_{\gamma})$ and $f(E_f, E_{\gamma})$ are shown for eight excitation regions. For all these deduced RSFs, we use one common level density in the evaluation of the
approximations (\ref{eq:Ei}) and (\ref{eq:Ef}). The data points of the experimental $P$ matrix cover only a certain region in $E_i$ and $E_{\gamma}$, making restrictions on the deduced RSFs. Thus, $f(E_i, E_{\gamma})$ is limited to 5.5 MeV~$ < E_i < S_n$ and 1.8 MeV~$ < E_{\gamma} < E_i$. The limits for $f(E_f, E_{\gamma})$ are $0 < E_f < S_n-1.8$ MeV and 1.8~MeV~$ < E_{\gamma} < S_n-E_f$. For a consistency check, we have tested that the average RSF for all $E_f$ energies equals the one for all $E_i$ energies (not shown here).

Figure~\ref{fig:rsfi} shows that the deduced RSF changes as a function of $E_i$, when low excitation energy is populated after the $\gamma$-emission. The lower, left panel, where two consecutive RSFs are plotted together, illustrates this. Here, the bumps seen at high $\gamma$-ray energies are the artifacts of the decay to specific isolated levels below 3 MeV of excitation. This is also the reason why apparently the fluctuations are typically a factor of $2-3$ in the panel where all deduced RSFs are plotted together (lower, right panel). In the plot of all deduced RSFs, we see a minimum at about $4-6$ MeV and some interesting structures at low $\gamma$-ray energies. These structures and the minimum are independent of the uncertainty in the log-slope of the level density, as shown in Fig.~\ref{fig:strength}.

The various $f(E_f, E_{\gamma})$ plots in Fig.~\ref{fig:rsff} are difficult to compare due to different limits appearing at both low and high $\gamma$-energies for the various $E_f$ regions. For example, the first three spectra do not reveal the region of low $\gamma$-energy enhancement because $E_{\gamma} > 5.5 {\rm ~MeV}-E_f > 3.7$~MeV. These spectra represent the decay to the $0^+$, $2^+$ and $4^+$ ground band states, respectively, where the experimental energy resolution makes some overlap between these states. In general the various spectra show strong fluctuations in the deduced RSFs when the $\gamma$ emission ends up at low excitation energy, typically $E_f < 3$~MeV. The panel with all deduced RSFs plotted together (lower, right panel) shows approximately the same scattering of data points as in Fig.~\ref{fig:rsfi}.

It is thus clear that the experimentally deduced RSFs for which states below $3$ MeV of excitation energy are involved in the $\gamma$ decay are very different. Figure~\ref{fig:counting} shows that this excitation region coincides with a region of few and well separated low-lying levels with specific structures. The bumps in the deduced RSFs are specific to the low energy level scheme, and the changes between the various deduced RSFs are large due to Porter-Thomas fluctuations.

In order to show the similarity of the deduced RSFs in the case of strongly suppressed Porter-Thomas fluctuations, we have compared the two uppermost excitation energy gates $E_i$ and $E_f$ in the lower, left panels of Figs.~\ref{fig:rsfi} and \ref{fig:rsff}, respectively. The deduced RSFs are here extracted for $\gamma$ decay between states in quasi-continuum, except for the data points with $E_{\gamma} > 7$~MeV in Fig.~\ref{fig:rsfi} where the final excitation energy is $< 3$~MeV. The deduced RSFs extracted from the quasi-continuum region behave similar to the deduced RSF from the all-over fit displayed in Fig.~\ref{fig:strength}. 

The good agreement between the deduced RSFs at higher energies is consistent with the expectation of suppressed Porter-Thomas fluctuations due to more initial and final levels in the evaluation of the $\gamma$ strength. These results strongly indicate that the concept of an RSF, which is independent on excitation energy, is valid already at relatively low excitation energies in $^{46}$Ti. At the same time, the results give us confidence that Oslo method works reasonably. The differences in deduced RSFs from lower energies indicate that the Porter-Thomas fluctuations are so poorly suppressed that it is difficult to predict the shape of the universal RSF. This is not very surprising and the differences in the deduced RSFs seem to be consistent with the expected fluctuations. 

In order to display the small differences in the deduced RSFs obtained from regions of higher level density, we have taken the four highest gates shown in Fig. 10 and only considered $\gamma$ energies up to 5.1 MeV. Thus, these statistically independent RSFs are evaluated in quasi-continuum with initial and final excitation energies of roughly $E_i = 8-10$ MeV and $E_f = 3-7$ MeV, respectively. The deduced RSFs are presented in the upper panel of Fig.~\ref{fig:fluct}.

For this data set we evaluated the relative deviations by
\begin{equation}
r_{ij} = \frac{f_{ij}-\left< f_i \right>}{\left< f_i \right> },
\end{equation}
where the index $i$ represents the $\gamma$ energy and $j$ is the initial excitation energy. The average strength at each $\gamma$ energy is estimated from the four individual RFSs
\begin{equation}
\left< f_i \right> = \frac{1}{n_j}\sum_{j}f_{ij}.
\end{equation}
In the lower panel of Fig.~\ref{fig:fluct} the $r_{ij}$ values are plotted showing the relative fluctuations from the mean value at each $\gamma$ energy.
The average ratio for all data points ($n_i=28$ and $n_j=4$) is taken as
\begin{equation}
r = \frac{1}{n_in_j}\sum_{ij}|r_{ij}|,
\end{equation}
giving $r \sim 6$\%. The differences in the deduced RSFs can easily be interpreted as remnants of the Porter-Thomas fluctuations. However, a quantitative estimate of the fluctuations is difficult to determine since the experimental statistical errors (see error bars) are also of the same order.

\section{Conclusions}
\label{sec:con}

The level density and radiative strength function for $^{46}$Ti have been determined using the Oslo method. Similar level density functions have been extracted from statistically independent data sets covering different excitation energies. This gives confidence to the Oslo method, since the disentanglement of the level density by Fermi's golden rule predicts one and only one unique level density, independent of the data set. 

The deduced RSF displays an enhancement at low $\gamma$-ray energy where we see a bump around 3 MeV and another structure at energies near 2 MeV. A similar enhancement has been seen in several other light mass nuclei and is still not accounted for by present theories.

A method to study the evolution of the deduced RSFs as a function of initial and final energy regions has been described. The deduced RSFs are found to display strong variations for different initial and final excitation energies if transitions to the lowest excitations are involved. The reason for the violent fluctuations of a factor of $2-3$ is that only a few isolated levels are present at low excitation energies with $E_f<3$~MeV. The differences in the RSFs obtained from a few number of transitions, that can be explained as a consequence of Porter-Thomas fluctuations of individual intensities, show that this energy region cannot be used for determination of the universal RSF. Even though, the deduced RSFs based on a restricted number of transitions still indicate that the decay is governed by a universal RSF.

However, the present work shows that it is possible to get more precise experimental information on the universal RSF. By imposing restrictions on the initial and final excitation energies, the RSFs for the decay between states in quasi-continuum can be extracted (i.e. for $E_f\gtrsim 3$~MeV). The results from this selected data set show that the decay is consistent with an RSF which is idependent of excitation energy within less than $\sim 6$~\% already at these relatively low excitations.
 
In summary, the observation of almost identical level densities and RSFs extracted from different data sets in quasi-continuum, indicates that the Oslo method works well. Provided that we use data from the quasi-continuum, a universal RSF in the light mass region of the $^{46}$Ti nucleus can be extracted. 

\acknowledgements
The authors wish to thank E.A.~Olsen and J.~Wikne for excellent experimental conditions. This work was supported by the Research Council of Norway (NFR).

\vfill

\begin{thebibliography}{99}

\bibitem{dirac}P.A.M. Dirac, Proc. R. Soc. London A {\bf 114}, 243-265 (1927).
\bibitem{fermi}E. Fermi, {\em Nuclear Physics} (University of Chicago Press, Chicago, 1950).
\bibitem{ENSDF}Data extracted using the NNDC On-Line Data Service from the ENSDF database. [http://www.nndcnbnl.gov./ensdf/]
\bibitem{dietrich}S.S. Dietrich and B.L. Berman, At. Data Nucl. Data Tables {\bf 38}, 199 (1988).
\bibitem{Schiller00} A.~Schiller, L. Bergholt, M.~Guttormsen, E. Melby, 
J.~Rekstad, S.~Siem, Nucl. Instrum. Methods Phys. Res. A {\bf 447}, 498 (2000).
\bibitem{CACTUS} M.~Guttormsen, A.~Atac, G.~L{\o}vh{\o}iden, S.~Messelt, T.~Rams{\o}y, J.~Rekstad, 
T.F.~Thorsteinsen, T.S.~Tveter, and Z.~Zelazny, Phys.\ Scr. \bf T 32\rm, 54 (1990).
\bibitem{gutt6} M.~Guttormsen, T.~S.~Tveter, L.~Bergholt, F.~Ingebretsen, and J.~Rekstad, 
Nucl.\ Instrum.\ Methods Phys.\ Res.\ A \bf 374\rm, 371 (1996).
\bibitem{Gut87} M.~Guttormsen, T.~Rams{\o}y, and J.~Rekstad, Nucl.\ Instrum.\
Methods Phys.\ Res.\ A \bf 255\rm, 518 (1987).

\bibitem{brink} D.~M.~Brink, Ph.D. thesis, Oxford University, 1955.
\bibitem{Andreas&Thoennessen} A.~Schiller and M.~Thoennessen, At. Data Nucl. Data Tables 93, 549 (2007).

\bibitem{PT} C.E. Porter and R.G. Thomas, Phys. Rev. {\bf 104}, 483 (1956).

\bibitem{Pb} N.U.H.~Syed, M.~Guttormsen, F.~Ingebretsen, A.~C.~Larsen, T.~L{\"o}nnroth, J.~Rekstad, 
A.~Schiller, S.~Siem, and A.~Voinov, Phys.\ Rev.\ C \bf 79\rm, 024316 (2009).
\bibitem{GC} A.~Gilbert and A.~G.~W.~Cameron, Can. J. Phys. {\bf 43}, 1446 (1965).


\bibitem{egidy2} T.~von Egidy and D.~Bucurescu, Phys.\ Rev.\ C \bf 72\rm, 044311 (2005); Phys.\ Rev.\ C \bf 73\rm, 049901(E) (2006).

\bibitem{goriely} S. Goriely, S. Hilaire, and A.J. Koning,, Phys.\ Rev.\ C \bf 78\rm, 064307 (2008).

\bibitem{Kristine10} Kristine Wikan, Master thesis in Physics, Department of Physics, University of Oslo, 
2010, http://www.duo.uio.no/publ/fysikk/2010/105030/Master.pdf 


\bibitem{Naeem09} N.U.H.~Syed, A.~C.~Larsen, A. \"{u}rger, M.~Guttormsen,
S. Harissopulos, M. Kmiecik, T. Konstantinopoulos, M.~Krti\u{c}ka, A. Lagoyannis, T. L\"{o}nnroth, K. Mazurek4, M. Norby, H. T. Nyhus, G. Perdikakis, S. Siem, and A. Spyrou, Phys.\ Rev.\ C \bf 80\rm, 044309 (2009).



\bibitem{Ohio04} Merico Salas-Bacci, Steven M. Grimes, Thomas N. Massey, Yannis Parpottas, Raymond T. Wheller, and James E. Oldendick, Phys.\ Rev.\ C \bf 70\rm, 024311 (2004).

\bibitem{BCS}J.~Bardeen, L.N.~Cooper, and J.R.~Schrieffer, Phys. Rev. {\bf 108}, 1175 (1957).
\bibitem{ko90} J. Kopecky and M. Uhl, Phys.~Rev.~C {\bf 41} 1941 (1990).
\bibitem{voin1} A. Voinov, M. Guttormsen, E. Melby, J. Rekstad,
A. Schiller, and S. Siem, Phys.\ Rev.\ C \bf 63\rm, 044313 (2001).
\bibitem{RIPL3} RIPL-3 Handbook for calculation of nuclear reaction, (2009); available at {\it http://www-nds.iaea.org/RIPL-3/}
\bibitem{Ishkhanov} B.S. Ishkhanov, I.M. Kapitonov, E.I. Lileeva, E.V. Shirokov, 
V.A. Erokhova, M.A. Elkin, A.V. Izotova, Moscow State Univ. Inst.~of Nucl.~Phys., Reports No.2002, p.27/711 (2002).
\bibitem{sc} A.C.~Larsen, M.~Guttormsen, R.~Chankova, F.~Ingebretsen, T.~L{\"o}nnroth, S.~Messelt, J.~Rekstad, 
A.~Schiller, S.~Siem, N.U.H.~Syed, and A.~Voinov, Phys.\ Rev.\ C \bf 76\rm, 044303 (2007).

\end{thebibliography}
\end{document}